\documentclass[twocolumn,amsmath,amssymb,superscriptaddress,pre]{revtex4-1}
\usepackage{amsmath,amscd,amsfonts,amssymb,color}
\usepackage{graphicx,amsfonts,epsf}

\usepackage{epsfig,graphicx}
\usepackage{dcolumn}
\usepackage{bm}
\usepackage{epsf}
\usepackage{amsmath,amsfonts}
\usepackage{mathtools}
\usepackage{amssymb}                                                                                                                                                                                                                                                                                                                                                                                                                     
\usepackage{color}
\usepackage[dvipsnames]{xcolor}
\usepackage[colorlinks=true,linkcolor=blue,urlcolor=blue,citecolor=blue,pdfusetitle]{hyperref}
\usepackage{hyperref}
\usepackage[dvipsnames]{xcolor}
\definecolor{redOA}{rgb}{0.8, 0.18, 0.1}

\newcommand{\bla}{bla\\bla\\bla\\bla\\bla}
\newcommand{\more}{bla\\more\\more}

\begin{document}

\title{Energy optimization of two-level quantum Otto machines
 } 


%

\author{Satnam Singh}
\email{satnamphysics@gmail.com \\satnamsingh@iisermohali.ac.in}
\affiliation{Department of Physical Sciences, Indian
Institute of Science Education \& 
Research (IISER) Mohali, Sector 81 SAS Nagar, 
Manauli PO 140306 Punjab India.}

\author{Obinna Abah}
\affiliation{Centre for Theoretical Atomic, Molecular and Optical Physics, School of Mathematics and Physics, Queen's University Belfast, Belfast BT7 1NN, United Kingdom}


\date{\today}


\begin{abstract}
We present the  spin quantum Otto  machine under different optimization criterion when function either as a heat engine or a refrigerator.    We examine the optimal performance of the heat engine and refrigerator depending on their efficiency, output power and maximum entropy production. For heat engine case, we obtain the expression for the upper and lower bounds efficiencies at maximum power and maximum ecological function. In addition, the spin quantum Otto refrigerator coefficient of performance is optimized for three different criterion -- cooling power, product of performance and power and ecological function.  We further study the dimensionless power loss to the cold reservoir when the machine is operating as a heat engine as well as its counterpart for the refrigerator case. We find that the maximum operation of the heat engine (refrigerator) cycle is when optimized with respect to hot (cold) reservoir frequency.

\end{abstract}

\maketitle 

\section{Introduction}
Heat engines and refrigerators are two main classes of the thermal machines that are important in our daily life. Heat engine converts the heat energy into the mechanical work, while the refrigerator absorb the heat energy from the lower temperature bath  and dump it into the higher temperature bath via the external work. 
Based on the second law of thermodynamics, the maximum efficiency of a traditional reversible and cyclic heat engine  pioneered by Sadi Carnot is $\eta_\text{C}\!=\!1-T_c/T_h$, where $T_c$ and $T_h$ are temperatures of the cold and hot reservoir respectively \cite{cal85}. The refrigerator is functioning as heat engine inverse and the associated maximum  coefficient of performance (COP) is  $\epsilon_C\!=\!T_c/(T_h - T_c)$ \cite{cal85}. However, the Carnot efficiency (COP) is reached only when the heat engine (refrigerator) is infinitely slowly operated to satisfy reversibility. For practical purposes in thermodynamics, engineering and biochemistry; it is important  to understand the thermodynamics optimization of irreversible thermal machines for best performance/efficiency \cite{Andresen2011}.
 
In particular, for heat engines, the efficiency at maximum power has been studied extensively and mainly characterized by the Curzon-Ahlborn efficiency $ \eta_{CA}\!=\!1-\sqrt{1-\eta_C}=\eta_C/2 + \eta_C^2/8+ \eta_C^3/16+ 5 \eta_C^4/128 + \mathcal{O}\left(\eta_C^5\right)$ \cite{Curzon1975AJP,EspositoPRL2010,Andresen2011,Deffner2018Entropy}. Although, the maximum power maximization counterpart of refrigerator is not straightforward, the COP at maximum cooling power of low-dissipation refrigerators is $\epsilon^\text{mp}\!=\!\epsilon_C/(2+\epsilon_C)$ \cite{Apertet:2013,Holubec:2020}. In addition, another meaningful figure of merit to characterize a refrigerator is the  product of the COP  and the cooling power of the refrigerator, the COP at maximum $\chi$ figure of merit, $\epsilon_\text{YC}\!=\!\sqrt{1+\epsilon_C} - 1 \!=\!\epsilon_C/2 - \epsilon_C^2/8 + \epsilon_C^3/16 - 5 \epsilon_C^4/128+ \mathcal{O}\left(\epsilon_C^5\right)$ \cite{Yan:1990,chen2001curzon,abah2016optimal}. Besides the maximum efficiency and maximum power criteria, Angulo-Brown  proposed the ecological optimization criterion of heat engines which take into account the trade-off between the high power output and the power loss due to entropy production, the Angulo-Brown efficiency $\eta_{AB}\!=\! 3 \eta _C/4 + \eta_C^2/32 + 3 \eta_C^3/128 + 37 \eta _C^4/2048 + \mathcal{O}\left(\eta_C^5\right)$ \cite{angulo1991ecological,ocampo2018thermodynamic}. 

Following the pioneering work of Scovil-Schulz-DuBois on a three-level maser heat engine \cite{scovil1959three}, there has been   progress in the development of quantum thermal machines \cite{bender2000quantum,humphrey2002reversible,Lin2004JPA,Kosloff:2010,Abah2012PRL,harbola2012quantum,thomas2012informative,goswami2013thermodynamics,wang2013efficiency,latifah2013quantum,sutantyo2015quantum,hofer2016quantum,Correa2016Entropy,dattagupta2017ericsson,yin2017optimal,chand2017single,singh2017feynman,newman2017performance,Roulet:2018,RojasGamboa:2018,Hewgill2018PRA,Oladimeji2019PhysE,singh2019three,chattopadhyay2019relativistic,singh2020multi,singh2020quantum,saputra2020quantum,barontini2019ultra,Myers:2020,Wiedmann:2020,Pena:2020}. These studies have investigated quantum version of the most  classical thermodynamics cycles, such as; Carnot, Otto, Diesel and Brayton. The working substances considered  are two-level atomic system \cite{wang2013efficiency}, harmonic oscillator \cite{Abah2012PRL,newman2017performance},  many-body systems \cite{Jaramillo:2016}, among others \cite{latifah2013quantum,sutantyo2015quantum,dattagupta2017ericsson,thomas2012informative,thomas2018quantum,hofer2016quantum,Li:2018}. 
Moreover, recent time, there has been tremendous success in miniaturization of thermal engines \cite{Rossnagel2016Science,Josefsson:2018,Horne:2020} and refrigerator \cite{Maslennikov2019NC} down to nanoscale as well as those operating in quantum regime \cite{Klatzow2019PRL,peterson2019experimental}.
 
 
 However, due to the increasing needs of energy consumption, resource availability, and environmental impact, the optimization of these real thermal engines/refrigerators are very desirable  \cite{long2016ecological,accikkalp2018performance}. Hernandez et. al. put forward a unified criterion for energy converters that is laying between those of maximum efficiency and maximum useful energy \cite{Hernandez:2001}. The ecological criterion for the heat engines  is $E_H\!=\! \dot{W} - T_c\, \dot{S}_\text{tot}$ while for the refrigerator, it is  $E_{R}\!=\!\dot{Q}_\text{out}-\epsilon_C T_h \dot{S}_\text{tot}$, where the dot (hereafter) is the time derivative with respect to the total cycle time, $W$ is the total work done,   $Q_\text{out}$ is the heat output, and $S_\text{tot}$ is the total entropy production \cite{Hernandez:2001}.

   In this work we study the optimal performance of the two-level Otto engine/refrigerator from the viewpoint of the efficiency, power and entropy production.   This model of the spin quantum heat engine is recently implemented  using the nuclear magnetic resonance setup \cite{peterson2019experimental}. Moreover, we study the power associated with optimal performance of the Otto cycle at different type of optimizations. Then, calculated the fractional power lost/dump of the engine/refrigerator cycle due to the  entropy production. 
   
   The remainder of the paper is organized as follows. In section \ref{model} we present the two-level atomic system thermodynamic quantities and the Otto cycle model. In Section \ref{OHE} we  present the analysis of Otto cycle when functioning as a heat engine. Then,  the optimal efficiencies are computed for two different optimization criterions, namely  the efficiency at maximum power (Section \ref{EMP}) and ecological function (Section \ref{EF}). We examine the refrigerator performance of Otto cycle for three different optimization in Section \ref{Otto-refrigerator} and in Section \ref{conc} we present our conclusions.

%


\section{Thermodynamics of two-level quantum Otto cycle} \label{model}
Let first discuss the thermodynamics of a quantum system.
The average internal energy of a quantum system with discrete energy levels is 
$ U\!=\!\sum_n E_n p_n$
where $E_n$ are the energy of the $n$-state/level and $p_n$ are the corresponding occupation probabilities. From an infinitesimally change in energy
\begin{equation}
 dU=\sum_n( E_n dp_n+ p_n dE_n)
 \label{int_energy}
\end{equation}
 we can distinguish the infinitesimal work done $dW\!=\!\sum_n dE_n p_n$ and the heat $dQ\!=\!\sum_n E_n dp_n$. Thus, Eq.~(\ref{int_energy}) can be seen as an expression of the first law of thermodynamics, $dU\!=\!dQ + dW$. 
 
 \begin{figure}
 \centering
 \includegraphics[scale=0.7,keepaspectratio=true]{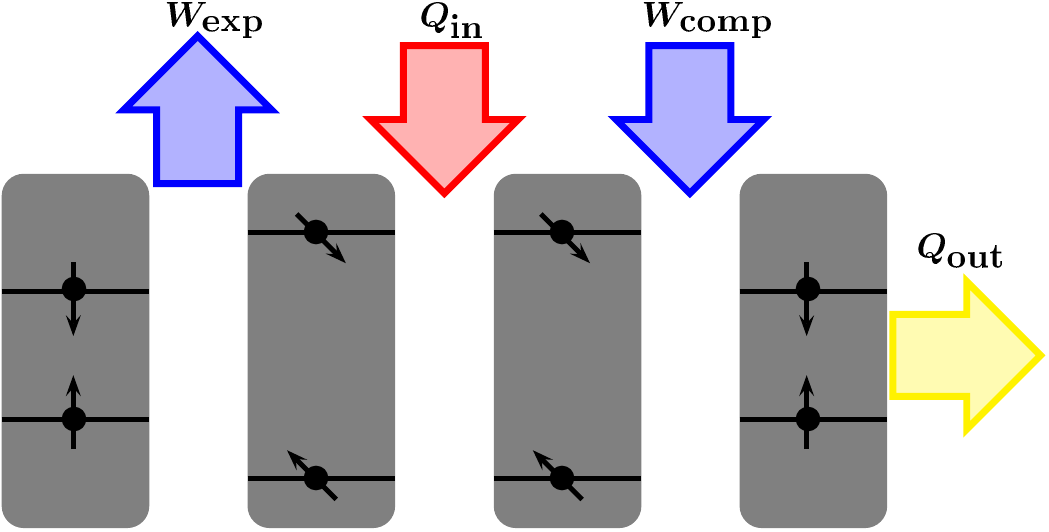}
 \caption{Schematic of a quantum Otto cycle constructed from the two-level system.  The thermodynamic cycle consists of two adiabatic processes ($W_\text{exp}$ and $W_\text{comp}$) and two isochoric processes ($Q_\text{in}$  and $Q_\text{out}$). } \label{fig:cycle}
\end{figure}
 We now consider a quantum Otto cycle whose working substance is a two-level system \cite{Lin2004JPA,Kosloff:2010,Abah2012PRL,Kosloff2017Entropy,Pena:2020}, see the pictorial representation in  the Fig.~(\ref{fig:cycle}). Specifically, for a two-level system described by the Hamiltonian $H\!=\!-\hbar \omega_t \sigma_z$, where $\hbar$ is the Planck constant, $\omega_t$ is the external controlled angular frequency and $\sigma_z$ is the $z$-component Pauli matrix. The associated occupation probabilities are given by $p^i_\pm\!=\!\exp(\pm\beta_i \hbar \omega_i)/Z_i$ and the partition function is $Z\!=\!\exp(\beta \hbar \omega_i)+\exp(-\beta_i \hbar \omega_i)$, where $i=c/h$ denotes the low/high angular frequency, $\beta_i=1/k_\text{B}T_i$ is the inverse temperature and $k_\text{B}$ is the Boltzmann constant.
The Otto cycle consists of two adiabatic branches where the external field $\omega_t$ varies with its energy-level structure and the two isochoric branches describes the working medium in contact with the cold/hot bath at constant control field $\omega$. The four-stroke stages of the cycle are ($\hbar\!=\!1$);\\ (i) \textit{adiabatic expansion} -- the two-level system initially prepared at frequency $\omega_c$ undergoes a unitary evolution  to reach a higher angular frequency $\omega_h>\omega_c$. The occupation probabilities for the two states remain unchanged according to the quantum adiabatic theorem \cite{mes99}. The work done during the expansion is given as
\begin{equation}
 W_\text{exp}=p_+^c(\omega_h-\omega_c)-p_-^c(\omega_h-\omega_c).
\end{equation}
(ii) \textit{isochoric heating} -- in this stage, the quantum system is coupled to the equilibrium hot thermal bath until it reaches the steady state at a constant angular frequency $\omega_h$. The work done during this process is zero and the corresponding heat input is given as
\begin{eqnarray}
 Q_\text{in}&=&-(\omega_h (p^h_+-p^c_-) - \omega_h(p^h_- - p^c_-)), \nonumber\\
 Q_\text{in}&=& \omega _h \left(\tanh \left(\beta _c \omega _c\right) - \tanh \left(\beta _h \omega _h\right)\right). 
\end{eqnarray}
(iii) \textit{adiabatic compression} -- the quantum system is isolated and the frequency varied from $\omega_h$ to $\omega_c$ at constant occupation probability. Similar to the expansion stage, no heat is added and  the work done during the adiabatic compression is 
\begin{equation}
 W_\text{comp} = p_+^h(\omega_c - \omega_h) - p_-^h(\omega_c - \omega_h).
\end{equation}
(iv) \textit{isochoric cooling step} -- the two-level quantum system  is coupled to the cold thermal bath temperature characterized by $\beta_c$.  The amount of heat discarded by the quantum system during this thermalization process reads
\begin{eqnarray}
 Q_\text{out} &=&-\left(\omega_c \left(p^c_+ - p^h_+\right) - \omega_c \left(p^c_- - p^h_-\right)\right), \nonumber\\
  &=& - \omega _c \left(\tanh \left(\beta _c \omega _c\right) - \tanh \left(\beta _h \omega _h\right)\right).
\end{eqnarray} 

For a complete cycle, the total work done becomes
\begin{eqnarray}
 W&=& -(W_\text{exp} + W_\text{comp}) \nonumber\\ &=& -\left(\omega _h - \omega _c\right) \left[\tanh\left(\beta _c \omega _c\right) - \tanh\left(\beta _h \omega _h\right)\right].
\end{eqnarray}
Thus, based on the first law of thermodynamics the amount of work $W$ produced by the engine or required by the refrigerator for any given cycle is 
\begin{align}
W = -(Q_c + Q_h). 
\label{1st_law}
\end{align}
In addition, an upper bound to the machine (engine/refrigerator) performance follows from the second law of thermodynamics, which states that the total entropy production of a cyclic thermal device is non-negative,
\begin{align}
S_\text{tot} = -\beta_h Q_h - \beta_c Q_c. 
\label{Stot}
\end{align}
In high-temperature limit, the total work done and heat input/output for a cycle can be written as
\begin{eqnarray}
 W&=&-\left(\omega _h - \omega _c\right) \left(\beta _c \omega_c - \beta_h \omega_h\right),\\
 Q_\text{in}&=& \omega _h \left(\beta_c \omega_c - \beta_h \omega_h\right), \\
   Q_\text{out}&=&- \omega_c \left(\beta_c \omega_c - \beta_h \omega_h\right).
\label{eqs:htemp}
\end{eqnarray}
In the rest of the paper, without lost generality, we will focus on the Otto cycle/machine operation at high-temperature limit.


\section{Two-level Otto heat engine}\label{OHE}
In this section, we will analyze the optimal performance of the two-level Otto heat engine using two different type of optimizations -- \textit{efficiency at maximum power} and \textit{ecological function}. Moreover, we study their fractional power loss and compare their maximum output power. 
For the cycle to function as heat engine, the total work done $W>0$. The efficiency of the quantum Otto heat engine is 
\begin{equation}
 \eta_O = -\frac{W}{Q_\text{in}}=1 - \frac{\omega_c}{\omega_h}\label{eq:eta}.
\end{equation}
The engine efficiency  depends on their initial and final frequencies. Based on the positivity of the total work done, $W\ge0$ which leads to the bound $1\le\omega_h/\omega_c\le \beta_c/\beta_h$, that is $\eta_C\!\ge\!\eta_O$. Alternatively, combining Eqs.~(\ref{1st_law}) and (\ref{Stot}), $\eta_C\!\ge\!\eta_O$. However, the maximum efficiency corresponds to zero output power, $P$ (i.e. total work done, $W$ per cycle time $\tau$) and occur when $\omega_c/\omega_h\!=\!T_c/T_h$.


\subsection{Efficiency at maximum power}\label{EMP}

\begin{figure}
 \centering
 \includegraphics[width=\columnwidth,keepaspectratio=true]{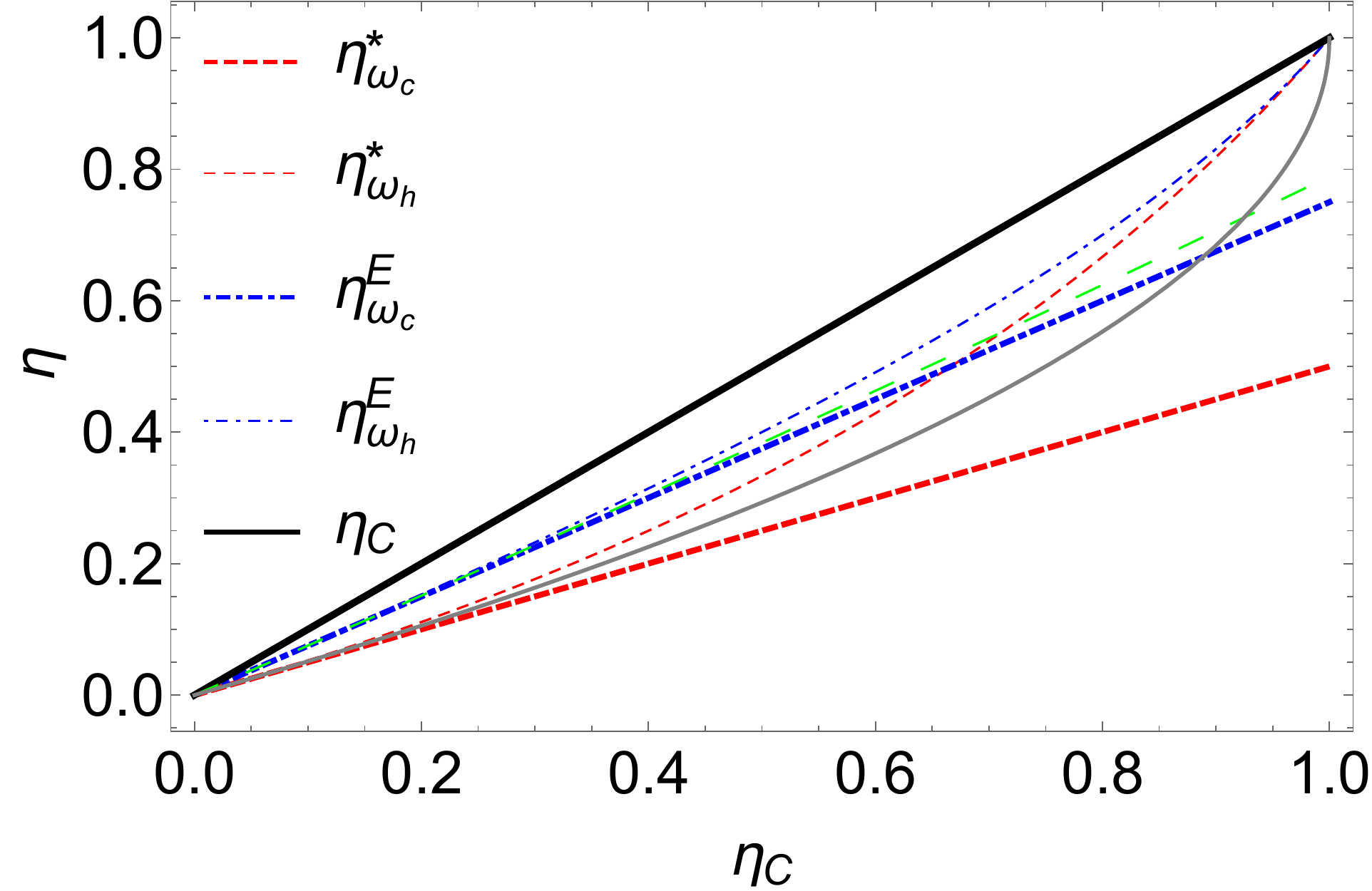}
 \caption{The Otto engine efficiency as a function of Carnot efficiency. The red (dashed) line is the efficiency at maximum work when optimized with $\omega_c$ (Eq.~\ref{eq:etaMP1}) while the red (light-dashed) line correspond to optimizing with $\omega_h$ (Eq.~\ref{eq:etaMP2}).  The blue (dotted dashed) line shows  the efficiency at $\omega_c$ maximum ecological function (Eq.~\ref{eq:eta-eco-omega1}) while the blue (light-dotted dashed) line is when the ecological function optimize with $\omega_h$ (Eq.~\ref{eq:eta-eco-omega2}). The black solid line is the Carnot efficiency $\eta_C$, the gray light solid line is the Curzon-Alhborn efficiency $\eta_{CA}$, while the green dashing line is the Angulo-Brown efficiency $\eta_{AB}$.}\label{fig:efficiency}
\end{figure}

Now we will optimize the power output of the heat engine cycle with respect to $\omega_h$ for fixed temperatures, cold frequency $\omega_c$ and cycle time.
The resulting optimal frequency ratio,  $\omega_c/\omega_h\!=\!2\beta_h/(\beta_c + \beta_h)$ with the corresponding efficiency and power are;
\begin{eqnarray}
 \eta_{\omega_h}^{\ast}&=&\frac{\eta_C}{2-\eta_C} \label{eq:etaMP2} \\
 P_{\omega_h}^\ast &=& \frac{\beta_c  \omega_c^2 \eta_C^2}{4 (1- \eta_C)\tau}.
\end{eqnarray}
Expanding  the efficiency  in terms of $\eta_C$, we have
\begin{equation}
  \eta_{\omega_h}^{\ast}=\frac{\eta_C}{2}+\frac{\eta_C^2}{4}+\frac{\eta_C^3}{8}+\frac{\eta_C^4}{16}+\mathcal{O}\left(\eta_C^5\right).
  \label{eq:etaEh}
\end{equation}
The smaller values of the $\eta_C$, the first term of Eq.~\ref{eq:etaEh} equals the $\eta_\text{CA}$ and  it is plotted in Fig.~\ref{fig:efficiency}. We observe that increasing values of $\eta_C$ gives optimal efficiency greater than the $\eta_\text{CA}$. At this point, it is worthy to mention that the  high-temperature limit efficiency at maximum power of a harmonic oscillator working medium is $\eta_\text{CA}$  \cite{Abah2012PRL}.

On the other hand, when we optimize the power output with respect to $\omega_c$ at fixed temperatures, $\omega_h$ and cycle time, the optimal frequency ratio is $\omega_c/\omega_h\!=\!(\beta_c+\beta_h)/2\beta_c$. The corresponding efficiency and power are
\begin{eqnarray}
 \eta^{\ast}_{\omega_c}&=&\eta_C/2 \label{eq:etaMP1},\\
 P^\ast &=& \frac{\beta_c\, \eta_C^2\, \omega_h^2}{4\, \tau}.
\end{eqnarray}
Equation (\ref{eq:etaMP1}) matches with the first term of $\eta_\text{CA}$ and illustrated in Fig.~\ref{fig:efficiency}. In general, the efficiency at maximum power is bounded as; $\eta^\ast_{\omega_c} \le \eta^\ast \le \eta^\ast_{\omega_h}$. We also note that the present  results almost agree well with the CA efficiency even for $\eta_C$ up to 0.3, at which the evident deviation of the present result from the CA efficiency starts to appear.

\subsection{Efficiency at maximum ecological function}\label{EF}
Let us consider the optimization of efficiency based on the ecological function with respect to the frequencies. In fixed cycle time, the ecological function defined as $E_{H}\!=\! \dot{W} - T_c\, \dot{S}_\text{tot}$ \cite{Hernandez:2001}, in high-temperature limit reads 
\begin{equation}
 E_{H}\!=\!- \left(\beta_c \omega_c - \beta_h \omega_h\right) \left( \omega_h\eta_C - 2\omega_h + 2 \omega_c\right).
 \label{eq:Eco-HE}
\end{equation}
We now optimize the ecological function, Eq.~(\ref{eq:Eco-HE}), with respect to $\omega_h$ to obtain the optimal frequency ratio as $\omega_c/\omega_h\!=\!2\beta_h(\eta_C - 2)/(\beta_c (\eta_C - 2) - 2\beta_h)$. The corresponding efficiency and power reads
\begin{eqnarray}
 \eta_{\omega_h}^{E}&=&\frac{\eta _C \left(2 \eta_C-3\right)}{3 \eta _C-4},\label{eq:eta-eco-omega2}\\
 P_{\omega_h}^{E}&=&\frac{3 \beta_c\,  \omega _c^2\, \eta_C^2 \left(2 \eta _c+1\right)}{4\, \tau \left(\eta_C + 2\right)^2 \left(\eta_C-1\right)}.
\end{eqnarray}
The expansion of the efficiency (Eq.~\ref{eq:eta-eco-omega2}) is
\begin{equation}
\eta_{\omega_h}^{E}=\frac{3 \eta_C}{4}+\frac{\eta_C^2}{16}+\frac{3 \eta_C^3}{64} + \frac{9 \eta_C^4}{256}+\mathcal{O}\left(\eta _C^5\right).
 \label{eq:eta-omega2-eco}
\end{equation}
From Eq.~(\ref{eq:eta-omega2-eco}), the first term is the same with the Angulo-Brown efficiency $\eta_{AB}$. It means that the for  the small values of the $\eta_C$, both efficiencies match with each other as illustrated in Fig.~\ref{fig:efficiency}.

Then optimizing the ecological  function with respect to $\omega_c$, the resulting optimal frequency ratio $\omega_c/\omega_h\!=\!\left[\beta_c\,(1-\eta_C/2) + \beta_h\right]/2\beta_c$. The corresponding efficiency and power are
\begin{eqnarray}
 \eta_{\omega_c}^{E}&=&\frac{3\eta_C}{4}, \label{eq:eta-eco-omega1} \\
 P_{\omega_c}^{E} &=& \frac{3\beta_c\, \eta_C^2\, \omega_h^2}{16\, \tau}.
\end{eqnarray}

From the Fig. \ref{fig:efficiency}, we can see that the efficiency at maximum ecological function is higher than the efficiency at maximum power.  The optimization with the $\omega_h$ gives the better results than the $\omega_c$. At the lower values of the $\eta_C$, the both  efficiencies at the the maximum ecological function matches with the Angulo-Brown efficiency $\eta_{AB}$, while both  the efficiencies at maximum power matches with the  Curzon-Ahlborn efficiency $\eta_{CA}$.  Likewise, the efficiency at maximum ecological function is bounded as; $\eta^E_{\omega_c} \le \eta^E \le \eta^E_{\omega_h}$.
We remark that the resulting power for the optimization of maximum power and ecological function with respect to $\omega_h$ is the same for $\beta_c\!=\!\omega_c\!=\!1$ while $P^\ast_{\omega_c} \ge P^E_{\omega_c}$ for $\beta_c\!=\!\omega_h\!=\!1$.

\subsection{Fractional power loss}
To better understand the different between the two efficiency optimization criterion in the Sections.~\ref{EMP} and \ref{EF}, we evaluate the ratio of power loss due to total entropy production (total entropy per unit time) to actual power output. Defining the power lost in terms of entropy production reads $P_\text{lost}\!=\!T_c \dot{S}_\text{tot}$ \cite{angulo1991ecological}, where $\dot{S}_\text{tot}\!=-\!\dot{Q}_c/T_c - \dot{Q}_h/T_h$. Using the definition of power $P\!=\!-\dot{W}\!=\!\dot{Q}_\text{in}+\dot{Q}_\text{out}$ and the efficiency $\eta\!=\!-W/Q_\text{in}\!=\!P/\dot{Q}_\text{in}$, the power loss reads
\begin{equation}
P_\text{loss} = \frac{P}{\eta} \left(\eta_C - \eta \right).
\end{equation}
Thus, the ratio of power loss to maximum power output can be written as
\begin{equation}
R = \frac{P_\text{loss}}{P} = \left(\frac{\eta_C}{\eta_\text{max}} - 1\right).
\label{Plost}
\end{equation}
Equation~(\ref{Plost}) quantifies the lost associated with the maximum efficiency $\eta_\text{max}$ for any optimization criterion. 

\begin{figure}
 \centering
 \includegraphics[scale=0.4,keepaspectratio=true]{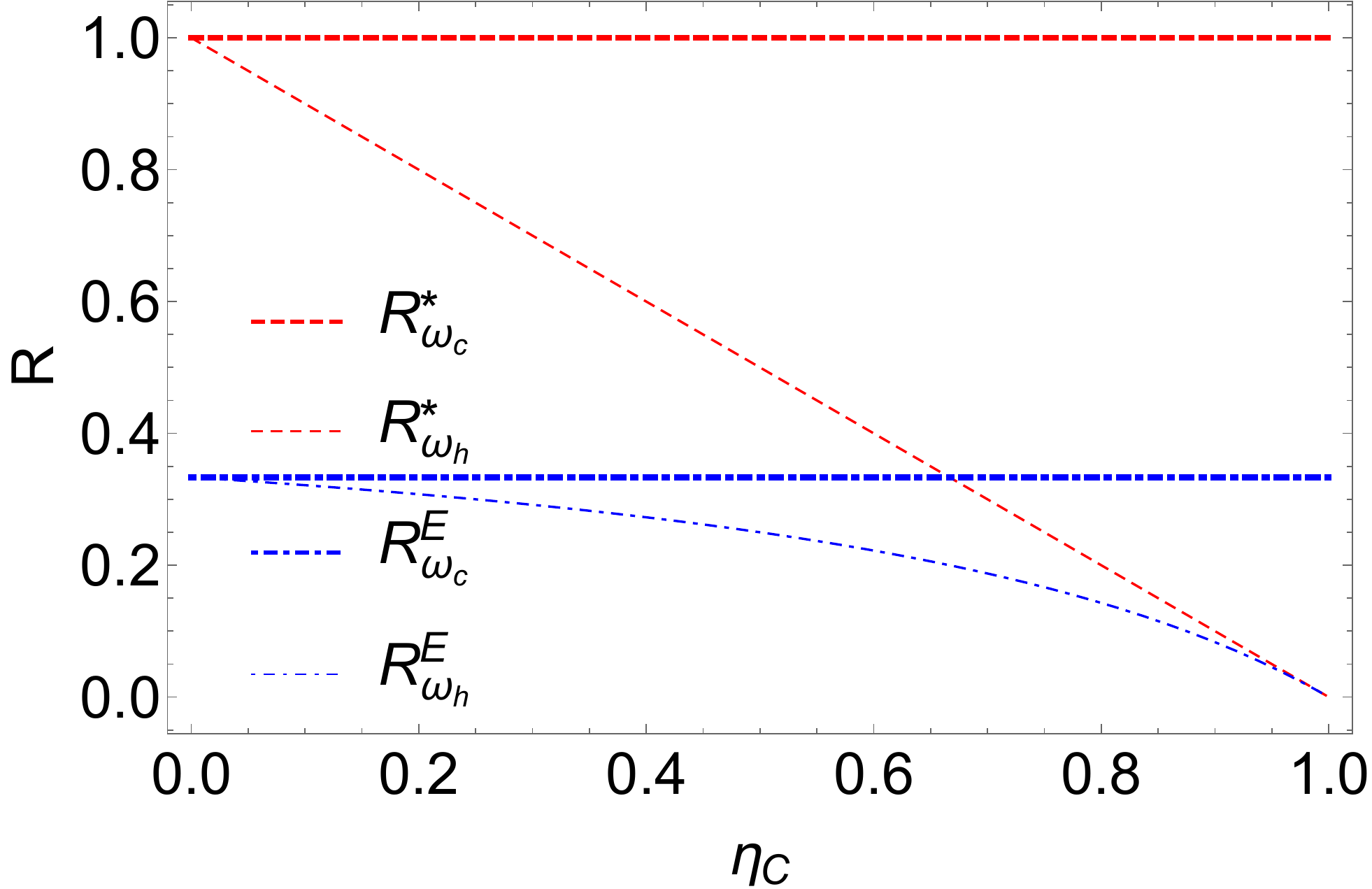}
 \caption{The dimensionless power loss with environmental temperature as a function of Carnot efficiency for different optimizations. The red (dashed) line corresponds to the maximum power optimization while the blue (dotted dashed) is the ecological function optimization.} \label{fig:loss}
\end{figure}

For the case of power optimization with respect to $\omega_h$, the fractional power loss for maximum power and ecological function respectively are
\begin{eqnarray}
 R_{\omega_h}^\ast\!=\!1-\eta_C, \hspace{1cm}
  R_{\omega_h}^{EF}\!=\!\frac{\eta_C - 1}{2\eta_C - 3}.
\end{eqnarray}
Similarly, the optimization of power with respect to $\omega_c$, we have
\begin{eqnarray}
 R_{\omega_c}^\ast\!=\!1 , \hspace{1cm}
  R_{\omega_c}^{EF}\!=\!1/3. 
\end{eqnarray}
Figure \ref{fig:loss} illustrate the fractional power loss as a function of temperature. We observe that the fractional power loss remains constant for optimization with  $\omega_c$, while it decreases with $\eta_C$, when  optimized with $\omega_h$.


In addition, the ratio of the heat engine maximum power and the power associated to maximum ecological function when optimised with $\omega_h$ reads $P^\ast_{\omega_h}/P^E_{\omega_h}\!=\!(\eta_C - 2)^2/(3-2\eta_C)$. On the other hand, the ratio when optimised with respect to $\omega_c$ is constant, i.e $P^\ast_{\omega_c}/P^E_{\omega_c}\!=\!4/3$.
Thus, it is clear that the ratio remains constant, when it is optimized with the $\omega_c$, while  the ratio decreases as we increase the $\eta_C$, when we optimize it with $\omega_h$

\section{Two-level Otto refrigerator}\label{Otto-refrigerator}
Here, we present the analysis of   the optimal performance of two level Otto refrigerator  for two different optimizations.  The main purpose of a refrigerator is to extract maximum possible of heat from the cold bath by performing a minimum amount of work.The Otto refrigerator  coefficient of performance (COP)  is defined as the ratio of  output heat $Q_\text{out}$ to  the total work done $W$ per cycle,
\begin{equation}
 \epsilon_\text{O} =\frac{Q_\text{out}}{W}=\frac{\omega_1}{\omega_2-\omega_1}.
 \label{eq:zeta}
\end{equation}
An Otto cycle functions as a refrigerator when the output heat is greater than zero. Based on the total entropy production for one complete cycle and the first law of thermodynamics, it can easily be shown that $\epsilon_\text{O}\!\le\!\epsilon_\text{C}$. Another important quantity describing a refrigerator is cooling power, $P_c$, defined as heat extracted from the cold bath per cycle over the cycle duration,
\begin{align}
P_c = Q_\text{out}/\tau.
\end{align}
For practical interest, we always have to find a compromise between cooling power and COP. 
However, it has been known that the optimization of refrigerator at maximum cooling power does not result to counterpart of efficiency at maximum power \cite{Apertet:2013}. Tom\'as et. al. proposed a unified optimization figure of merit as product of COP and cooling power of a refrigerator \cite{Tomas:2012}. In what follows, we study refrigerator performance for three different optimization criterion.

\subsection{Optimization of the  cooling power}
Here, we find the maximum COP for a given cooling power. Maximizing the cooling power at constant cycle time with respect to  $\omega_c$ gives the optimal cold frequency $\omega^\ast\!=\!\beta_h\omega_h/2\beta_c$ and the corresponding performance quantities as,
\begin{eqnarray}
 \epsilon^\text{mp}\!&=&\!\frac{\epsilon_c}{2 + \epsilon_c} \label{eq:zeta-mp}, \hspace{1cm}\\ P_c^\text{mp}\!&=&\!\frac{\beta_h \epsilon_\text{C} \omega_h^2}{4\tau(1+\epsilon_\text{C})}\label{Pc-mp}.
\end{eqnarray}
The Taylor's expansion of the COP is 
\begin{equation}
 \epsilon^\text{mp}\!=\! \frac{\epsilon_c}{2}-\frac{\epsilon_c^2}{4}+\frac{\epsilon_c^3}{8}-\frac{\epsilon_c^4}{16}+\mathcal{O}\left(\epsilon_c^5\right)
\end{equation}
The $ \epsilon^\text{mp}$, Eq. (\ref{eq:zeta-mp}), is the same as the result obtained recently for Carnot-type low-dissipation refrigerators in the reversible limit \cite{Apertet:2013,Holubec:2020}. We see that the lower values of  performance at the maximum cooling power is similar to the result of Yan-Chen \cite{Yan:1990,chen2001curzon}. We remark that optimization with respect to $\omega_h$ leads to no physical results.

\begin{figure}
 \centering
 \includegraphics[width=\columnwidth,keepaspectratio=true]{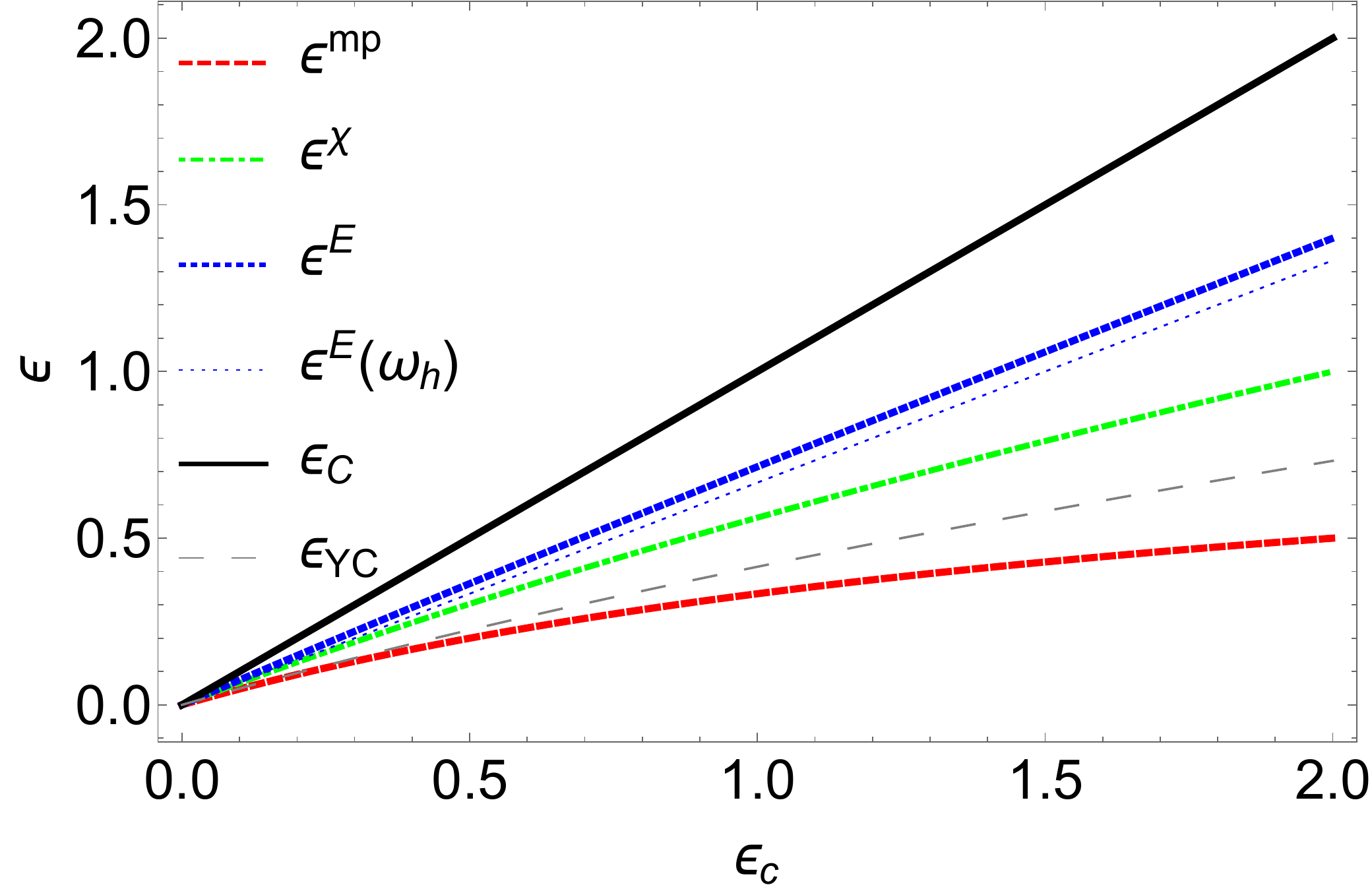}
 \caption{Coefficient of performance at different maximum optimization for the Otto refrigeration cycle as a function of the Carnot coefficient of performance $\epsilon_\text{C}$. The red (dashed) line shows the maximum cooling power case (Eq. \ref{eq:zeta-mp}), the green (dotted-dashed) line shows the $\chi$ figure of merit case (Eq.~\ref{COP-chi}), the blue (dotted) line shows the ecological function  case (Eq.~\ref{COP-EF}) when maximised with $\omega_c$, while the blue (dotted) tiny line is when ecological function is maximised with respect to  $\omega_h$. The black solid line is the Carnot coefficient of performance while the gray (large dashing) line is the Yan-Chen coefficient of performance, $\epsilon_\text{YC}$.} \label{fig:zeta}
\end{figure}
\subsection{Performance at maximum $\chi$ figure of merit}
Now let consider the unified figure of merit $\chi\!=\!\epsilon\,Q_\text{out}/\tau$
defined as the product of the coefficient of performance $\epsilon$ and the cooling power of the refrigerator \cite{Tomas:2012,abah2016optimal}.
Optimizing the $\chi$ with respect to $\omega_c$  in the high-temperature limit gives the optimal frequency ratio,
\begin{align}
\frac{\omega_c}{\omega_h}\!=\!\frac{-\sqrt{9 \beta_c^2-10 \beta_c \beta_h + \beta_h^2}+3 \beta _c+\beta_h}{4 \beta_c}.
\end{align}
The associated COP at maximum $\chi$ and cooling power are;
\begin{eqnarray}
\epsilon^\chi\!=\! 1/2 (-3 + \sqrt{(9 + 8 \epsilon_\text{C})}), \label{COP-chi}\\
P_c^\chi\!=\! \beta_h\frac{\left[(3+2\epsilon_\text{C}) \sqrt{9+8\epsilon_\text{C}} - (9+10\epsilon_\text{C}) \right] \omega_h^2}{8\tau \epsilon_\text{C}(1+\epsilon_\text{C})}\label{Pc-chi}.
\end{eqnarray}
Similar to optimization at maximum cooling power, the COP can be express in the form of Yan-Chen COP \cite{Yan:1990}.


\subsection{COP at maximum ecological function}
We now evaluate the COP at maximum ecological function, $E_{R}\!=\!\dot{Q}_\text{out}-\epsilon_C T_h \dot{S}_\text{tot}$. The  ecological function of the two-level Otto refrigeration cycle  in high temperature limit becomes
\begin{equation}
  E_R=\frac{\left(\beta _h \omega_h - \beta_c \omega_c\right) \left(\beta_c \omega_c \epsilon_\text{C}+\beta_h \left(\omega_c - \omega_h \epsilon_\text{C}\right)\right)}{\beta_h}.
\label{eq:eco-refri}
\end{equation}
First,   optimizing the ecological function of the Otto refrigerator with respect to $\omega_h$ at fixed temperatures, $\omega_c$ and cycle time, we have $ \omega_h\!=\!(\beta _h \omega_c+2 \beta_c \omega_c \epsilon_\text{C})/(2 \beta_h \epsilon_\text{C})$ and the resulting COP and cooling power are
\begin{equation}
   \epsilon_{\omega_h}^{E}=2 \epsilon_\text{C}/3 \label{eq:zeta-omega2-eco}, \hspace{1cm} P_c^E(\omega_h) = \frac{\beta_h \omega_c^2}{2 \tau \epsilon_\text{C}}.
\end{equation}
Alternatively, optimizing with respect to $\omega_c$,  the optimal frequency $ \omega_c\!=\!\beta _h \omega _h \left(\beta _h+2 \beta _c \epsilon_C\right)/(2 \beta _c \left(\beta _h+\beta _c \epsilon_C\right))$ with the COP at maximum ecological function and cooling power read
 \begin{eqnarray}
  \epsilon^{E}(\omega_c)\!&=&\!\frac{\epsilon_\text{C} \left(2 \epsilon_\text{C} + 3\right)}{3 \epsilon_\text{C}+4},
   \label{COP-EF} \hspace{0.3cm}\\
 P_c^E(\omega_c)\!&=&\!\frac{\beta_h \epsilon_\text{C} (2 \epsilon_\text{C} + 3) \omega_h^2}{4 \tau (\epsilon_\text{C} + 1) (\epsilon_\text{C} +2)^2}\label{Pc-EF}.
 \end{eqnarray}
 We remark that the resulting COP for maximization with respect to $\omega_c$,  $ \epsilon^{E}(\omega_c)$ is slightly greater than $ \epsilon_{\omega_h}^{E}$.
 
\begin{figure}
 \centering
 \includegraphics[width=\columnwidth,keepaspectratio=true]{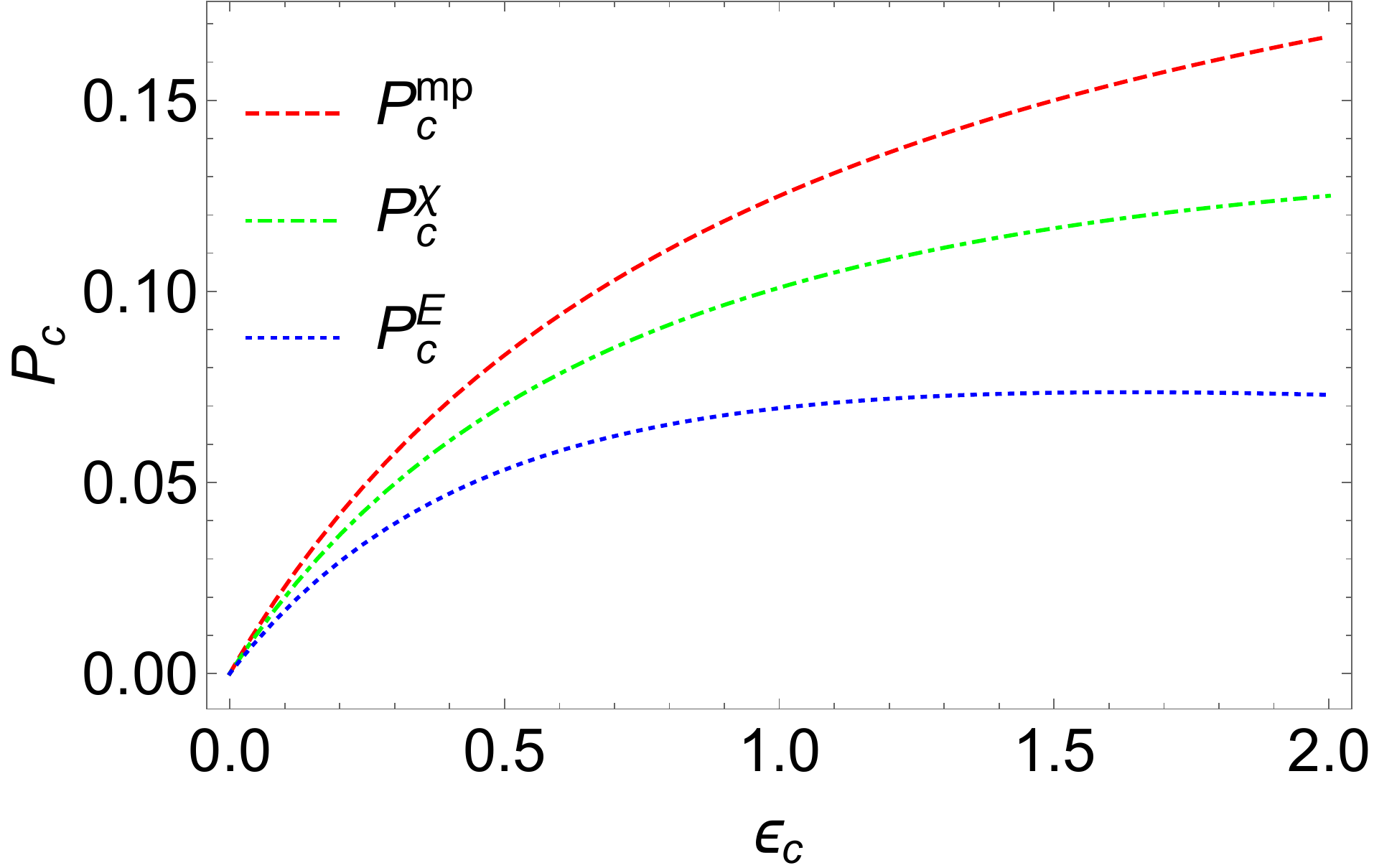}
 \caption{Cooling power at different maximum optimization for the Otto refrigeration cycle as a function of the Carnot coefficient of performance $\epsilon_\text{C}$. The red (dashed) line shows the maximum cooling power case (Eq. \ref{Pc-mp}), the green (dotted-dashed) line shows the $\chi$ figure of merit case (Eq.~\ref{Pc-chi}), while the blue (dotted) line shows the ecological function  case (Eq.~\ref{Pc-EF}) when maximised with $\omega_c$. The cooling power is in the unit $\beta_h\omega_h/\tau$. } \label{Pcooling}
\end{figure}

Figure \ref{fig:zeta} shows  the COP at maximum cooling power (Eq.~\ref{eq:zeta-mp}, red dashed line), the COP at maximum $\chi$ figure of merit (Eq.~\ref{COP-chi}, green dotted dashed line) and COP at ecological function maximization (Eq.~\ref{COP-EF}, blue dotted line).  We observe that the COP all concides for large temperature difference (small $\epsilon_\text{C}$). The COP at maximum ecological function and the $\chi$ figure of merit are greater than the Yan-Chen COP, $\epsilon_{YC}$. Thus,  the COP are related as; $\epsilon_\text{C}^\text{mp} \le \epsilon_\text{C}^\chi \le \epsilon_\text{C}^E$. In Fig.~\ref{Pcooling}, we present the corresponding maximum cooling power for the different optimization. Their behaviour is inversely related to the maximum COP illustrated in Fig.~\ref{fig:zeta}.

\begin{figure}
 \centering
 \includegraphics[width=\columnwidth,keepaspectratio=true]{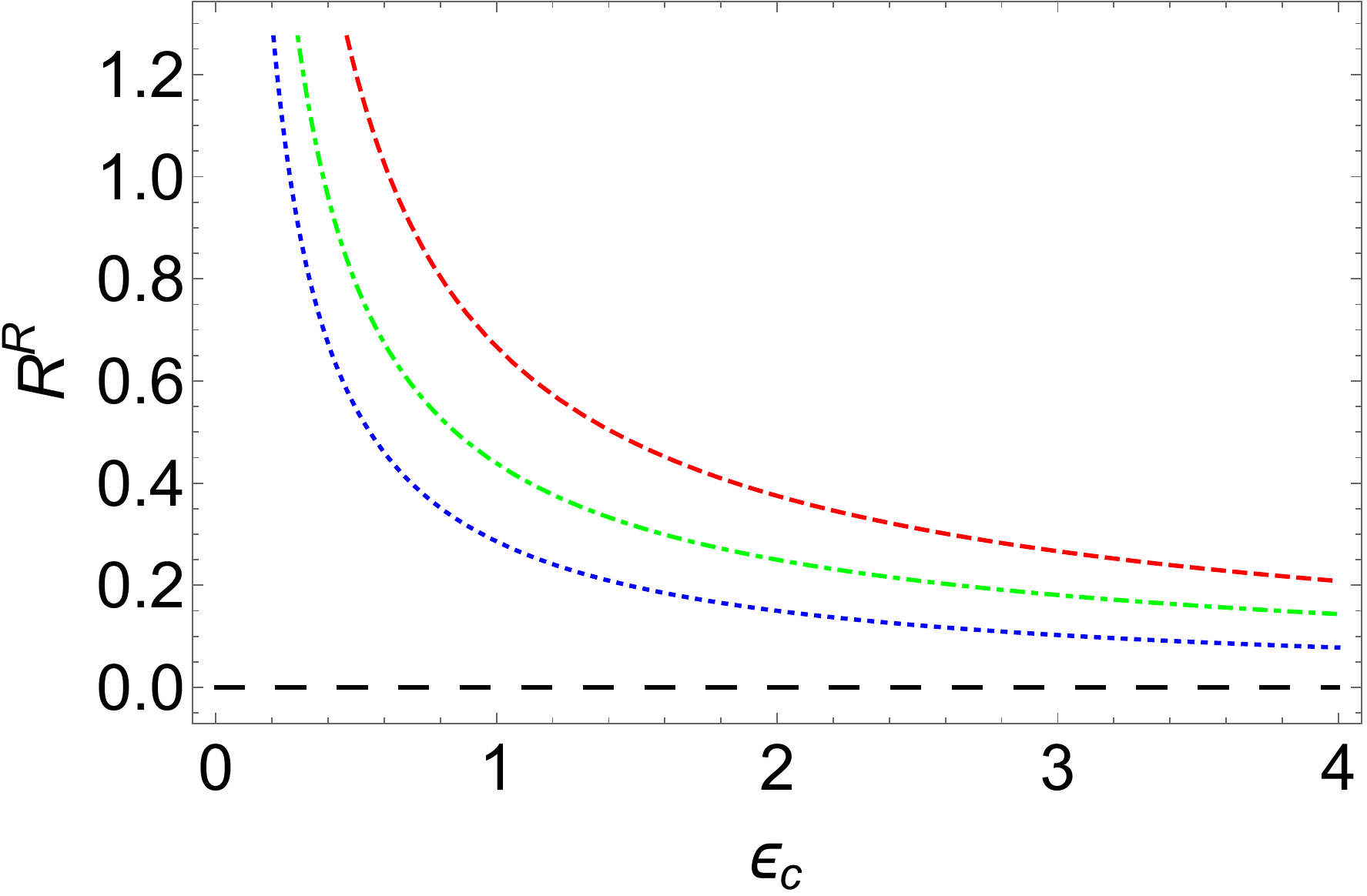}
 \caption{The fractional cooling power loss at  different optimization as function of Carnot coefficient of performance. The red (dashed) line shows the maximum cooling power case (Eq.~\ref{pc-loss-mp}), the green (dotted-dashed) line shows the $\chi$ figure of merit case (Eq.~\ref{pc-loss-chi}), while the blue (dotted) line shows the ecological function  case (Eq.~\ref{pc-loss-E}). The large dashed black line is a guide to the eyes on point zero.} \label{fig:loss-cooling-power}
\end{figure} 

\subsection{Fractional cooling power dump }
Here, in analogy to the fractional power loss of heat engine, we consider the dimensionless cooling power \textit{dump} during a complete refrigeration cycle. Let us define the power dump due to the entropy generation in the hot reservoir as 
\begin{equation}
P^R_\text{dump} \!=\!T_h \Delta \dot{S}_\text{tot} =T_h \left[-\frac{\dot{Q}_h}{T_h} - \frac{\dot{Q}_c}{T_c}\right],
\end{equation}
where the environment temperature is equal to the hot temperature.


Employing the definitions, Carnot COP $\epsilon_C\!=\!T_c/(T_h-T_c)\!=\!\beta_h/(\beta_c -\beta_h)$ and Otto COP $\epsilon\!=\!Q_c/W$, we get
\begin{equation}
P^R_\text{dump} = \dot{Q}_c\left[\frac{1}{\epsilon} - \frac{1}{\epsilon_C}\right].
\end{equation}
Thus, the dimensionless cooling power of any Otto refrigerator becomes
\begin{equation}
\mathcal{R}^R=\frac{P^R_\text{dump}}{P} = \left[\frac{1}{\epsilon} - \frac{1}{\epsilon_C}\right].
\label{Ref_loss}
\end{equation}
In the first order approximation of $\epsilon_C$, Eq. (\ref{Ref_loss}) can be re-written as
\begin{equation}
\mathcal{R}^R=\frac{1}{\epsilon_C^2} \left(\epsilon_C - \epsilon_\text{max}\right).
\end{equation}
where $\epsilon_\text{max}$ is the resulting COP for a given optimization criterion.
Thus, the cooling power loss associated with the COP at maximum cooling power, $\chi$ figure of merit and the ecological function when optimized with $\omega_c$  respectively are;
\begin{eqnarray}
\mathcal{R}^R(mp) &=& \frac{1 + \epsilon_\text{C}}{2\epsilon_C + \epsilon_C^2}, \label{pc-loss-mp}\\
\mathcal{P}^R(\chi) &=& \frac{3+ 2\epsilon_C - \sqrt{9 + 8\,\epsilon_C} }{2 \epsilon_C^2}, \label{pc-loss-chi}\\
\mathcal{R}^R(E) &=& \frac{1+\epsilon_\text{C}}{4\epsilon_C + 3\epsilon_\text{C}^2}. \label{pc-loss-E}
\end{eqnarray}

Figure \ref{fig:loss-cooling-power} present the dimensionless cooling power loss as a function of Carnot COP for different optimization protocol. 
It is clear that the power loss decreases as we increase the $\epsilon_\text{C}$ and the case   of maximum ecological function gives lowest fractional power loss. In addition, the ecological function optimization with respect to $\omega_h$ yields a  dimensional cooling power loss $\mathcal{R}^R(E_{\omega_h})\!=\!1/(3\, \epsilon_C)$  that is higher than $\mathcal{R}^R(E)$ but the same in high values  of $\epsilon_\text{C}$.




\section{Conclusion}\label{conc}
We have studied the quantum Otto cycle whose working medium is a two-level system, first when functioning as a heat engine and later as a refrigerator. For one complete cycle, the two-level system alternate between two (hot and cold) thermal reservoirs by varying their angular frequency from $\omega_c$ to $\omega_h$. For heat engine, we analyze the optimal efficiency at maximum power as well as ecological function and find that the optimal efficiency at maximum ecological function is always greater than the maximum power case. Then, optimizing with respect to $\omega_h$ yields more efficiency than the case of $\omega_c$ for a particular optimization.  In addition, we calculated the dimensionless power loss to the cold environment due to change in entropy per unit time and observe that while the optimization with respect to $\omega_c$ is constant, the optimization with $\omega_h$ depends on Carnot efficiency $\eta_C$. Moreover, the amount of power loss to environment is minimal for the case of ecological function than the maximum power.

On the other hand, we have studied the performance of Otto refrigeration cycle for optimal cooling power, $\chi$-function figure of merit and ecological function. The ecological function optimization gives highest COP while the cooling power optimization leads to the lowest values. However, the amount of power dump into the hot environment which decreases with the COP is more  for the maximum cooling power than the ecological function scenario and always finite.
Finally, we conclude that the heat engine cycle remains more beneficial when it is optimized with the $\omega_h$, while refrigeration cycle remains more beneficial when it is optimized with the $\omega_c$.
\section{Acknowledgement}
We are thankful to  Colin Benjamin and Varinder Singh for their valuable comments. OA acknowledge the support by  the Royal Commission for the Exhibition of 1851.

\bibliography{engine,QHE-reference}
\bibliographystyle{apsrev4-1}

\end{document}